\documentclass[doublecol]{epl2}
\usepackage{siunitx}

\title{Social distancing and the future of pedestrian dynamics}

\shorttitle{Social distancing and the future of pedestrian dynamics}

\author{Mohcine Chraibi\inst{1} \and Andreas Schadschneider\inst{2,3} \and  Antoine Tordeux\inst{4} }
\shortauthor{M. Chraibi, A. Schadschneider, A. Tordeux}

\institute{                    
    \inst{1} Institute of Advanced Simulation, Forschungszentrum J\"ulich GmbH, 52425 J\"ulich, Germany, \email{m.chraibi@fz-juelich.de}\\
  \inst{2} Institut f\"ur Theoretische Physik, Universit\"at zu K\"oln, 50937 K\"oln, Germany, \email{as@thp.uni-koeln.de} \\
  \inst{3} Institut f\"ur Physikdidaktik, Universit\"at zu K\"oln, 50931 K\"oln, Germany,\\
  \inst{4} School of Mechanical Engineering and Safety Engineering,
University of Wuppertal, 42285 Wuppertal, Germany \email{tordeux@uni-wuppertal.de}
}

\abstract{Many countries have introduced social distancing rules during the COVID-19 pandemic in order to reduce spreading of the virus in public spaces. This has inspired research on pedestrian dynamics in more ways than just one. Many studies have been performed, not only to determine the effectiveness of such measures, but also in order to understand its effects on crowd dynamics in general. 
In this Perspective article, we reflect on the insights derived from these investigations and their relevance for future advancements in the field of pedestrian dynamics. 
This includes impacts on safety regulations and the potential for new theoretical and experimental approaches which may hold relevance even beyond applied restrictions.
}

\begin{document}

\maketitle


\section{Introduction}

Pedestrian dynamics is an interdisciplinary field that studies the movement of people in different scenarios, such as buildings, streets, and public spaces. Understanding how pedestrians move is crucial for designing safe and efficient infrastructures, as well as for ensuring the well-being of individuals in crowded places. In the last 25 years, physicists have made many important contributions to the understanding of pedestrian streams. One important aspect is the interplay between experiments and modeling. Laboratory experiments allow researchers to study pedestrian behavior under controlled conditions which provides valuable insights into the basic mechanisms underlying their movement. For instance, experiments in simple scenarios such as bottlenecks or corridors have revealed how pedestrians form queues and interact with each other to avoid collisions. 

Beyond the remaining open questions in the field, due to the COVID-19 pandemic new challenges appeared, e.g.\ understanding the effects of social distancing rules. 
This Perspective article aims to provide a concise summary of current research in pedestrian dynamics, with a particular emphasis on the latest developments driven by the COVID-19 pandemic. 
Furthermore, future research directions will be discussed as well as the lessons social distancing has taught us in general about pedestrian motion.

One of the major new challenges is to understand how social distancing measures affect pedestrian flow and interactions, as well as how individual behaviors influence the spreading of the virus. 
We will limit ourselves to discussing research on pedestrian movement under social distancing restrictions and its impact on pedestrian behavior in public spaces. 
Of course, there exist other important aspects that are relevant for public health measures and efforts to control the spread of viruses like COVID-19.
We will not go into much detail concerning the spreading of the virus as it is not directly related to the understanding of pedestrian motion. 
However, most models discussed could be combined with a model of virus spreading to get more quantitative insights into the spreading of infections.


\section{State-of-the-Art before COVID-19}

Our understanding of pedestrian and crowd dynamics has been improved enormously in the last 25 years, not at least through contributions from physics. 
It shows several surprising collective phenomena that can be understood and modeled using ideas borrowed from the physics of many-body systems. 
Arguably, the most important innovation put forward by physicists is a close interplay between experimental and theoretical work. 
In recent years, many large-scale experiments with pedestrians have been performed under controlled (laboratory) conditions which provided accurate data. 
These results have then been used to improve and calibrate modeling approaches that are useful, e.g., for planning and improving safety at mass events with thousands of people.
However, although much progress has been made, certain aspects are not sufficiently understood yet, especially in high-density situations where experiments are not feasible for ethical reasons.

\subsection{Model approaches}

A large variety of crowd motion models have been proposed (for reviews, see e.g.\
\cite{Martinez-Gil2017,SchadschneiderCSTZ18,ChraibiTSS18,FelicianiBook,CorbettaToschi}) which can be classified according to different criteria. 
Most models used are microscopic, i.e., they consider the motion of each individual. 
One of the most important classes here are cellular automata models where all variables (space, time, and state) are discrete. 
The motion of a pedestrians is determined by stochastic rules where the transition probabilities to neighbouring positions depend on the current state of the pedestrian considered and her immediate neighbourhood. 
Another class are acceleration- or force-based models which are continuous in space and time. 
Pedestrians are typically represented by circular or elliptical disks interacting via forces that can be physical (e.g.\ friction) or non-physical (so-called social forces). 
The dynamics is then determined by the coupled Newton equations of all particles. 
Often, a hard-core exclusion between the particles is assumed, sometimes with a ``soft surface'' representing elastic deformations. 
The exclusion results from the superposition of distance-dependent repulsive forces, that
can take different forms, such as  short-range exponential forces (as seen in~\cite{Totzeck2020}), algebraic forces (as in~\cite{Yu2005,Chraibi2010}), or even partly linear forces~\cite{Nakayama2005,CordesST20}. 

 In recent years, it has been realized that force-based models suffer from intrinsic problems (
 e.g., violations of the exclusion principle) which often are related to dominating inertia effects, especially at high density levels~\cite{Chraibi2011,Sticco2020,CordesST20}. 
 Therefore, first-order velocity-based models 
 have gained increasing popularity. 
 These models inherently assume pedestrian hard-core exclusion among pedestrians, effectively disregarding inertia in the system~\cite{Maury2011,Tordeux2016}.
The primary interactions of a pedestrian with the environment (e.g., other pedestrians or infrastructure) occur along the direction of motion. 
This characteristic renders pedestrian dynamics highly anisotropic.
Anisotropy in pedestrian models can be introduced, for instance, through various approaches, such as a vision field factor~\cite{Yu2005,Chraibi2010}, specifying a preferred avoidance direction~\cite{Totzeck2020}, or utilizing a speed function that considers only the distance to the next pedestrian or obstacle in the direction of motion~\cite{Maury2011,Tordeux2016}.
While such models are realistic for ``normal'' situations, they face challenges when considering social distancing requirements. 
Social distancing necessitates maintaining safe distances not only in the direction of motion, but also in the space around each pedestrian. 
Models designed for social distancing will consequently require improved isotropic features, including lateral and backward interaction mechanisms, along with extended anticipation capabilities for crossing pedestrians.

\subsection{Empirical investigations}

Highly controlled laboratory experiments are crucial to gain insights into the complex inner workings of pedestrian movement and have significantly contributed to our understanding of crowd dynamics. 
They provide a valuable tool for controlling variables that would be impossible to control in real-world settings. By isolating specific factors that influence pedestrian dynamics, researchers can develop effective policies and urban planning strategies to ensure the safety and efficiency of pedestrian traffic. These experiments typically employ simple scenarios that replicate realistic pedestrian flows, such as bottlenecks or counter-flows in corridors. In many cases the experiments start with relatively homogeneous groups, such as students, to establish baseline behaviors before exploring the influence of heterogeneity. Laboratory experiments offer the advantage of being conducted under controlled conditions, including factors like lighting and camera position, which significantly simplify the recording and measurement of pedestrian behavior and thus the analysis process.

While physical quantities such as density, speed, and flow can be reliably studied in laboratory experiments, factors like social norms, cultural background, and individual characteristics that influence pedestrian dynamics are more difficult to capture. Consequently, experiments are often conducted in different countries to test for cross-cultural variations in pedestrian behavior. For instance, in~\cite{Chattaraj2009}, Indian and German participants were compared in a laboratory experiment to investigate the effect of cultural background on pedestrian movement. The study found that it significantly affects pedestrian behavior, with Indian participants moving faster in high-density situations. Similarly, studies on single-file movement have been conducted in various countries including Germany, India, China, and Palestine, revealing that the fundamental diagram (density-speed relationship) varies depending on factors such as culture, age, and gender~\cite{subaih2019gender,Paetzke2022}.

An unresolved challenge is the study of crowd movement at high densities, which cannot be easily achieved in controlled experiments due to safety and ethical concerns. In these experiments, the trajectories of individual pedestrians are typically recorded using high-precision video analysis. This process can be performed semi-automatically, allowing for the analysis of experiments with over 1000 participants in a reasonable amount of time. From the recorded trajectories of pedestrians, macroscopic quantities such as flow and density can be extracted. 

\begin{figure}[!ht]
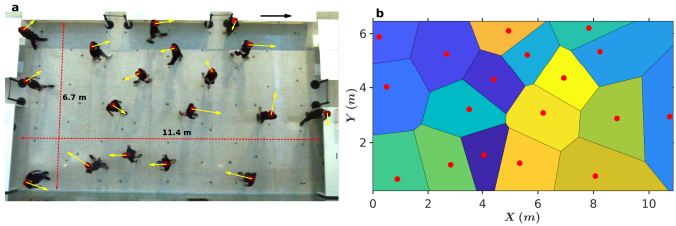

\onefigure[width=\columnwidth]{Figures/Fig1.eps}
\caption{The experiment depicted in the image shows 18 pedestrians walking at a slow pace while maintaining a prescribed safety distance of \qty{1.5}{\meter}.
  The Voronoi diagram in (b) is calculated from (a) with edge restrictions of the finite system. Each pedestrian's Voronoi cell is colored differently, allowing the calculation of the area $A_i$ and corresponding individual density $\rho_i$. Source: \cite{Echeverria-Huarte2021}.}
\label{fig.1}
\end{figure}

However, the interpretation of these quantities can be complicated due to the use of different density definitions. The simplest definition involves counting the number of individuals within a specific area. Due to the physical size of pedestrians, determining whether a person is considered to be within the area or not can be difficult. Furthermore, large fluctuations in the density can occur over time as individuals enter or leave the area, especially if the area is small. To circumvent these issues, a Voronoi tessellation-based definition for density has been proposed~\cite{Steffen2010a}. 
This approach considers each person as the center of a Voronoi cell, i.e.\ a polygon containing the area that is closer to that person than any other individual (see Fig.~\ref{fig.1}, right panel). The inverse of the area of the Voronoi cell can then be interpreted as a local density. This definition is less sensitive to fluctuations caused by persons entering or leaving the considered area and provides a more precise representation of density.

In summary, laboratory experiments offer a valuable means of studying pedestrian dynamics and identifying factors that influence movement. However, it is important to acknowledge their limitations in capturing all aspects of real-world behavior. A comprehensive understanding of pedestrian behavior can be achieved by combining laboratory experiments with field studies and data analysis from real-world settings. Future research should strive to develop advanced laboratory experiments that can account for the complexities of real-world behavior by incorporating both physical and social factors and combining their benefits with real-world measurements~\cite{Corbetta2014a}.


\section{Social distancing}

The requirement for social distancing has clearly impacted the behavior in public spaces. This has been investigated in wide range of studies, including both experimental investigations and modeling endeavors. 
Specifically, the research has predominantly focused on two key aspects: 1) evaluating compliance with social distancing rules, and 2) investigating the influence of social distancing on crowd properties, such as flow rates, densities, and jamming probabilities.

\subsection{Experimental and empirical work} 

\begin{figure*}
    \centering
    \includegraphics[width=.9\textwidth]{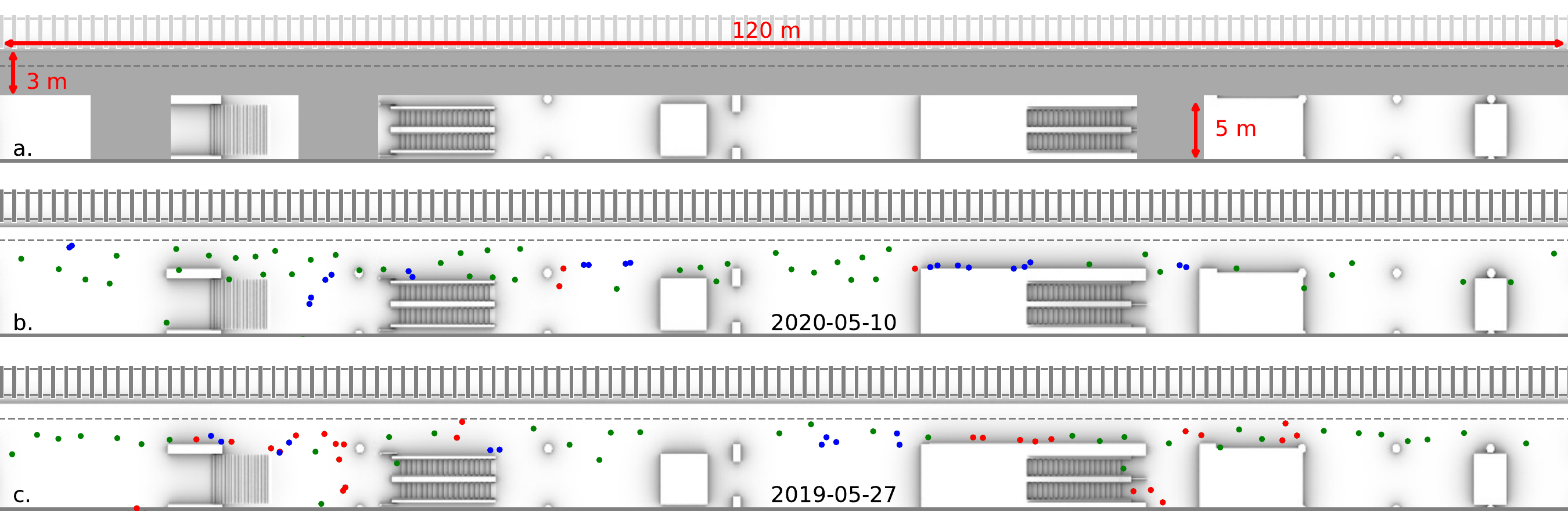}
    \caption{Floorplan of Platform 3 at Utrecht Central Station (NL) showing the monitored area in gray. (b) Sample of 75 passengers waiting for a train on May 10, 2020, during the Covid-19 pandemic, with green indicating those adhering to physical distancing, blue for family-group members, and red for distance offenders. Only 3 out of 75 people violate the rules. (c) Similar sample on May 27, 2019, before the pandemic, showing one-third standing closer than \qty{1.5}{meter} to someone else. Source: \cite{Pouw2020}}
    \label{fig:pouw2020}
\end{figure*}

An important part of experimental research has concentrated on examining the compliance of individuals with social distancing requirements and identifying the factors that influence their adherence to the prescribed distances. The goal of these studies is to gain a better understanding of the determinants and challenges related to the ability to maintain the necessary physical distancing. 
For instance in~\cite{Pouw2020}, a highly efficient and precise method for analyzing, in real time, distance-time encounter patterns in crowds  was developed. Pedestrian tracking data  collected at a platform in Utrecht Central Station in the Netherlands (Fig.~\ref{fig:pouw2020}) revealed a high correlation between contact times and passenger numbers, which may indicate a decrease in adherence to social distancing rules and an increase in the difficulty of complying with regulations.

Similar observations were revealed by CCTV footage of public interactions in inner-city Amsterdam \cite{Hoeben2021}. Violations of the \qty{1.5}{\meter} distance guidelines in the weeks prior to, during, and after the guidelines were implemented during the COVID-19 pandemic were analyzed. Initially, compliance with the guidelines was  high, but began to decline quickly. The study also revealed a strong correlation between the number of individuals observed on the street and violations of the \qty{1.5}{\meter} distance guidelines. 

In \cite{Schneider2021}, observation data were utilized to predict the likelihood of physical distancing compliance on recreational trails across six U.S.\ states. The study aimed to understand the influence of site, visitor, and time-related factors on compliance with physical distancing recommendations. 
Documenting more than 10,000 trail user encounters, it was found that wider trails, smaller groups, and signage were associated with greater compliance with physical distancing recommendations. The study also revealed that compliance with physical distancing tends to decrease over time, but innovative and dynamic signage could help sustain compliance. 

In addition to empirical observations, several experiments have been performed under laboratory conditions. 
In~\cite{Echeverria-Huarte2022} experimental findings on pedestrian evacuations through a narrow door while adhering to prescribed safety distances of either \qty{1.5}{\meter} or \qty{2}{\meter} were reported. 
The absence of flow interruptions or clogs, which are typically observed under normal conditions, results in an increase in flow with pedestrian velocity. However, this contrasts with the `faster-is-slower' effect observed in narrow bottlenecks~\cite{Garcimartin2014}. 
Evacuation efficiency improves when the prescribed physical distance is reduced, as this shortens the time between consecutive pedestrians exiting. The flow rate remains independent of crowd size, suggesting a negligible effect of psychological crowd pressure on the evacuation process. The seemingly contradictory roles of crowd size can be explained by considering that, in these special conditions of physical distancing, the flow is only regulated at the exit and is determined solely by pedestrian walking speed and interpersonal distance. 
Thus, increasing walking speed and reducing the safety distance positively affect the flow rate, while the size of the evacuated crowd has a negligible effect on the evacuation process.

The relationship between pedestrian speed and density when individuals maintain a certain personal social distance were investigated experimentally in~\cite{Echeverria-Huarte2021}. Participants were asked to walk at different walking speeds in a rectangular room while maintaining a personal social distance of either \qty{1.5}{\meter} or \qty{2}{\meter}. The study found that the speed-density relation for a \qty{1.5}{\meter} personal social distance was similar to the one for a \qty{2}{\meter} distance, with the most significant difference being a small variation caused by different global densities for slow walking speeds when the personal social distance was \qty{1.5}{\meter}. This suggests that individual pedestrian dynamics is dominated by collision avoidance when the personal social distance is reduced. The study also found that the speed-density relation is primarily determined by motivation (in terms of choosing a fast or slow walking speed) for low densities, while the global density has a noticeable effect on the speed-density relation when the walking speed is low and the personal social distance is large. This is likely because reducing conflicts among pedestrians in terms of close approaches provides individuals with enough time to decide their direction while taking into account not only their immediate neighbors, but also their perception of the rest of the room. 

The same experiments were 
analyzed in~\cite{Echeverria-Huarte2021a} to investigate the impact of pedestrian density, walking speed, and prescribed safety distance on interpersonal distance.
The results suggest that a density of no more than 0.16 pedestrians per square meter is required to ensure a minimum interpersonal distance of \qty{1}{\meter}. Even at this density, over 50\% of interpersonal distances fall below \qty{2}{\meter}. As pedestrian density and walking speed increase, the percentage of safety distance violations further increases.
The study also finds that a prescribed safety distance of \qty{2}{\meter} is more effective than \qty{1.5}{\meter} in reducing the percentage of time that pedestrians are closer than \qty{1}{\meter}. 
Overall, the results suggest that pedestrian density is the primary factor determining interpersonal distance in moving crowds and that lower densities and longer safety distances are preferable to ensure safety in dense crowds.

A study conducted at the University of Canterbury~\cite{Marshall2020,Ronchi2021} 
aimed to examine the evacuation behavior from a room under different social distancing restrictions, specifically considering scenarios with open and closed doors. 
The baseline scenario involved no social distancing rules, and three additional scenarios were tested with different social distancing rules. 
The average total travel time through the exit increased significantly when social distancing rules were implemented, as interpersonal distance increased. The comparison of the two  scenarios with social distancing of \qty{2}{\meter} with open and closed doors revealed that the average total travel time decreased when the door was not held open. However, the scatter in the data indicated that social distancing rules were not strictly adhered to, possibly due to the limited experience of the participants with social distancing.
The average interpersonal distance increased as the experiment progressed, indicating that people can get used to social distancing and better judge interpersonal distances. 

%

The study~\cite{Lu2021} aimed to examine the behavior of pedestrians under different social distancing measures by conducting single-file experiments with two types of prescribed distances of \qty{1}{\meter} and \qty{2}{\meter}. The comparison with pre-pandemic experiments reveals that while social distancing measures do encourage pedestrians to maintain a greater distance, violations still occur, which is in accordance with previous experiments. 
Despite the presence of a typical stop-and-go wave in experiments, the increase in social distancing caused pedestrian stopping behavior to occur at lower density and shortened the density range of the transition from free flow to jammed flow. 
This leads to the emergence of several research questions such as the level of compliance among pedestrians under different social distancing measures, whether pedestrian stopping behavior occurs earlier under these measures, and whether the typical stop-and-go wave still exists.
The comparison with results for two-dimensional movement~\cite{Echeverria-Huarte2021a} shows
that pedestrians are more likely to comply with the prescribed safety distance and stop less in single-file movement.
This is reasonable since in single-file movement, pedestrians only need to estimate the distance to the person in front. 
In two-dimensional movement, however, the distance to everyone around them must be estimated, which makes it more difficult to comply with the prescribed safety distance.

In summary, the empirical studies on compliance with distance restrictions have revealed \emph{quantitatively} that adherence to these guidelines is often limited. 
One contributing factor is the challenge individuals face in accurately assessing and maintaining the required distances. 
However, there is evidence to suggest that with sufficient training and practice, people can improve their ability to comply with these distance restrictions. 
Furthermore, the studies have shown that increasing densities have a negative impact on compliance. 
Future studies should quantitatively investigate the relationship between density and compliance and explore strategies for improving compliance rates.


\subsection{Modeling}

During the pandemic, 
pedestrian simulation models were used to investigate the impact of social distancing on pedestrian behavior and disease spreading. These studies typically utilize variations of established models, like social force models, agent-based models, and particle-based simulations. 
Also new metrics and methodologies for evaluating the effectiveness of social distancing measures in various environments were introduced and tested.

\begin{figure}[!ht]
\onefigure[width=.7\columnwidth]{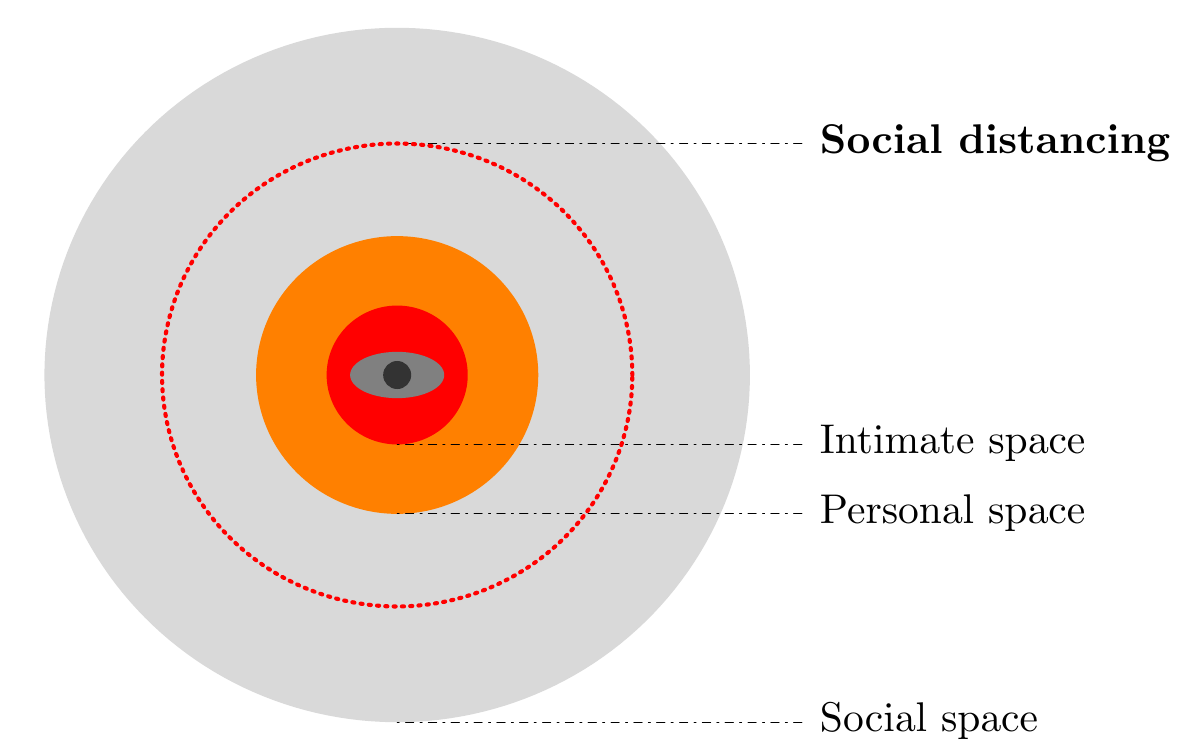}
\caption{Edward T. Hall's interpersonal distances of man \cite{hall1966hidden}: intimate distance (\qty{0.6}{m}), personal distance (\qty{1.2}{m}) and social distance (\qty{3}{m}) in comparison with the usual social distancing of \qty{2}{m} (dotted circle).}
\label{fig:SocDist}
\end{figure}

As described above, several experimental studies clearly demonstrated that pedestrians can not perfectly estimate distances which is an important factor for the observed violations. One straightforward approach to implement social distancing in existing models would be to adjust the size of pedestrians. 
For instance, it has been proposed to represent each pedestrian by a rigid disk with a radius larger than their shoulder width to ensure the desired spacing~\cite{MykoniatisAS21,ZhangSUFK22}.
This idea resonates with the interpersonal distance zones theorized by Edward T. Hall~\cite{hall1966hidden}, consisting of intimate distance, personal distance, and social distance, which are approximately 0.6, 1.2, and 3 meters, respectively (see Fig.~\ref{fig:SocDist}). 
A pedestrian can be modelled using the intimate disk, while the social distance is in between the personal and the social disc.

However, while using rigid disk to represent each pedestrian including social distancing  restricts violations of the social distancing rule, it can cause clogging, particularly in bottleneck situations such as exits. 
Therefore, finding a balance between enforcing the rule and allowing for practical considerations remains an ongoing challenge.

To avoid such problems and the need to introduce new models, other studies have tried to capture the social distancing behavior by adjusting model parameters. In \cite{Alam2022}, a social force model was used to simulate passenger flows at an international airport during the COVID-19 pandemic. 
Several other studies used a similar methodology, see, e.g., \cite{TsukanovSFB21,alam2022pedestrian,YangJKYLY22}. 
Varying the repulsive forces among agents,  results in relatively few instances where pedestrians come within 2 meters of each other. This observation aligns with the general trends observed in empirical data~\cite{Echeverria-Huarte2021a}.

The studies \cite{Cui2022,Sajjadi2021} follow a similar methodology in investigating the impact of social distancing on the movement and infection risk of individuals, by using the social force model integrated with virus transmission models.
These studies found that social distancing does not always reduce exposure time or high-risk encounters, when the proportion of people wearing masks incorrectly was 50\% and in some setups due to the limitation of the site.

In~\cite{Espitia2022}, the impact of social distancing on pedestrian dynamics was investigated, focusing on the parameters of the repulsive force in the social force model, particularly the range parameter used to indirectly model social distancing. 
A noticeable decrease in capacity was observed in simulations under SD-restrictions using a circular setup. 
The maximum density also experiences a decline from its initial value, which was used for calibrating the model parameters.

A similar approach was followed in~\cite{Xu2020} by employing numerical simulations to investigate the effectiveness of measures aimed at reducing contact among customers in a supermarket using a velocity-based model~\cite{Xu2019}. 
To quantitatively evaluate the degree of contact between customers, an index based on the interpersonal distance is defined. 
The study reveals that areas with intersecting pathways in supermarkets have a significantly higher rates of customer contact. 
Therefore, a potential strategy for reducing such interactions could involve altering the store layout to minimize these crossing areas. This adjustment, combined with continued enforcement of social distancing rules, could further enhance preventative measures within these spaces.

Particle-based simulations are used in \cite{Kramer2021} to study the impact of social distancing on pedestrian flows. It is shown that social distancing can be thought of as an increase in the internal length scale of the flow, specifically the radius within which pedestrians repel each other (but with the size of the pedestrians held constant). 
By changing the range parameter of the repulsive force that pedestrians experience it is found that social distancing reduces the density in a corridor, with a greater effect observed at higher densities. 

In~\cite{Mayr2021}, new parameter values for the Optimal Steps Model are determined in order to simulate social distancing. The methodology involved deriving regression formulas for two cases, one where contacts had to be avoided completely and another where a short violation of the social distance was allowed. 
The study found a linear correlation between personal space and desired social distance, indicating that as individuals strive for greater social distance, their personal space proportionally increases. This expansion in personal space was observed to extend evacuation times, occasionally leading to temporary clogging where agents become momentarily immobile. Interestingly, an increase in social distance was found to increase the likelihood of such clogging, a finding that seems to contrast with~\cite{Echeverria-Huarte2022} where increased pedestrian velocity led to a rise in flow without any observed clogs.

In a different approach, a modification of the social force model was suggested by replacing the exponential repulsive force with a force derived from Lennard-Jones potential~\cite{SI2021}. The ``social distance'' is here identified with the distance at which the particle-particle potential energy vanishes which corresponds to a  
socially distanced crowd that is packed in a confined space. The use of the Lennard-Jones potential is deemed problematic and unstable in situations where social distancing restrictions were violated. 
However, the authors did not discuss this issue in their paper. By comparison with the exponential social force model and experimental data from~\cite{Echeverria-Huarte2021a}, it is argued that the model can match the PDF functions of social distancing.
In~\cite{Ronchi2020}, a new methodology, named EXPOSED, was presented which aims to estimate occupant exposure in confined spaces. The method is model-agnostic: it is designed to work with any microscopic crowd model capable of simulating pedestrian movement and providing location data over time. Similarly, in the study~\cite{Usman2020} a new metric called the Social Distancing Index (SDI) was introduced which can be used to evaluate the effectiveness of a space layout in maintaining safe social distances. This approach is comparable to the approach suggested in~\cite{Xu2020}.
\section{Future perspectives}
\label{sec-future}

The COVID-19 pandemic and the associated restrictions on personal contact, including social distancing rules, have presented new challenges for the field of pedestrian dynamics. These challenges have inspired a surge of research, both experimental and theoretical. 
However, the urgency of the situation led to an unprecedented pace of research and publication. 
While this rapid response was necessary given the circumstances, it also meant that many studies were conducted concurrently. 
As a result, researchers often had limited opportunities to build upon the findings of earlier work before embarking on their own investigations. 
This is not a critique of the researchers, who were responding to an urgent global crisis, but rather an observation about the unique circumstances for research during this period.

Despite these challenges, through these investigations, researchers have gained a better understanding of the relationship between social distancing rules and pedestrian dynamics. 
They have explored topics such as adherence to prescribed distances, changes in flow patterns, density fluctuations, and the impact of different scenarios on overall crowd behavior. Furthermore, novel metrics and methodologies were introduced that allow to assess the effectiveness of social distancing measures in different environments.

For future pandemics, we are probably much better prepared. Based on the experiences made during the COVID-19 pandemic, we know which measures are necessary to reduce virus spreading in crowds and, most importantly, how to implement them in a way that is acceptable for pedestrians which is essential to improve compliance.  
Counter-flows or crossing streams should be avoided as they can easily lead to inevitable violations of the social distancing rules. 
Instead uni-directional streams are preferable, but not always easy to realize. 
In the future, safety regulations should be updated to accommodate those findings and the layout of new buildings should be improved to simplify the realization of such measures.

Finally, these investigations have allowed a new perspective on crowd dynamics in general. 
It has been suggested that social distancing can be used as a kind of magnification glass for pedestrian interactions.
In general, conducting laboratory experiments with crowds at high densities poses challenges due to safety and ethical concerns. 
Additionally, accurately detecting and tracking individual pedestrians can be challenging, particularly when occlusion effects occur. 
However, one advantage of conducting experiments under social distancing rules is that these challenges are mitigated. By ensuring a prescribed distance between individuals, the detection and tracking of pedestrians becomes more manageable. This allows for a clearer observation and analysis of their behavior within the given constraints. 
Therefore, conducting experiments under social distancing rules provides a unique opportunity to investigate pedestrian dynamics in a controlled and safer environment. 
It allows researchers to study the effects of social distancing on crowd behavior without the complexities associated with high-density scenarios. 
However, it remains to be seen in how far the behaviour under real high densities and high densities in social distancing scenarios are comparable. 
In complement, virtual reality technology could provide a 
controlled environment for studying pedestrian behavior in various scenarios. It has the potential to bridge the gap between laboratory experiments and field studies, allowing for more accurate data collection while maintaining experimental control. 
By combining empirical research, theoretical modeling, and advanced experimental techniques, researchers can work towards creating comprehensive models that enhance our understanding of pedestrian dynamics and improve safety and efficiency under social distancing requirements.


\acknowledgments
We thank the authors of \cite{Echeverria-Huarte2021} and \cite{Pouw2020} for permission to reproduce figures~\ref{fig.1} and \ref{fig:pouw2020}.
MC and AT acknowledge the Franco–German research project MADRAS funded 
by the Agence Nationale de la Recherche (ANR
), grant number ANR-20-CE92-0033, and 
by the Deutsche Forschungsgemeinschaft (DFG
), grant number 446168800.



\end{document}